\documentclass[sigconf, nonacm]{acmart}
\renewcommand\footnotetextcopyrightpermission[1]{}

\usepackage{tabularx}
\usepackage{subcaption}
\usepackage{graphicx}
\AtBeginDocument{%
  }

\begin{document}

\title[Adaptive Affective Chatbots: Control, Automation, and Transparency]
{Designing Adaptive Affective Chatbots for Online Learning: Trade-offs Between User Control, Automation, and Transparency}

\thanks{This preprint presents an extended version of the study published in the SIGCSE Virtual 2026 proceedings \cite{jung2026adaptive}.}


\author{Jay Y. Jung}
\email{jy.jung@gatech.edu}
\affiliation{%
  \institution{Georgia Institute of Technology}
  \city{Atlanta}
  \state{GA}
  \country{United States}}
  
\begin{abstract}
Online students often face emotional challenges such as frustration, isolation, and fluctuating motivation, which can hinder sustained engagement in online learning. While affective chatbots have shown promise in healthcare and wellness, most learning support chatbots focus primarily on cognitive assistance with limited attention to learners' emotional and motivational experiences. In this work, we investigate how adaptive affective chatbot support can be designed for online learning through an iterative design process.

Through a mixed-methods needfinding study ($n=38$), we identified substantial variation in how learners respond to affective support, with no single strategy universally preferred. This led us to develop three adaptive design alternatives—user-controlled, AI-driven, and hybrid co-controlled adaptation—and examine key trade-offs between automation, transparency, and user control. A final evaluation ($n=25$) showed that 88\% of participants favored the adaptive hybrid design over a non-adaptive alternative, highlighting the value of balancing transparency, user control, and interaction simplicity. We discuss design implications for adaptive affective chatbot systems in emotionally challenging online learning contexts.
\end{abstract}





\maketitle

\section{Introduction}

Recent advances in artificial intelligence (AI) have accelerated the integration of chatbot-based systems into educational environments, opening new possibilities for AI-driven learning support. In educational contexts, chatbots serve diverse roles—as teaching agents, peer agents, and motivational agents—to support self-directed learning, provide explanations, and assist learners during challenging academic tasks \cite{hobert2019say, chhibber2019using}. In the design of effective AI-driven educational tools, prior work emphasize engagement, trust, and emotional effects as important factors \cite{rapp2021human}.

In online learning environments (OLEs), chatbots are recognized to promote engagement with the potential to enhance learning outcomes \cite{robinson2008new}. Recent evidence has further shown that chatbots can reduce cognitive load and improve learning performance through formative feedback \cite{yin2024using}. However, while prior studies have highlighted the importance of emotional engagement in self-regulated learning context with AI tutors \cite{jin2023supporting}, there has been limited research on how chatbots can provide not only cognitive assistance but also affective support particularly for online learners. Specifically, existing studies have not examined how affective chatbot interactions should be integrated for the emotional and motivational challenges that are often encountered in online learning environments. Although emotionally intelligent chatbots have shown promise in domains such as health and wellness, their potential in educational settings remains underexplored.

To explore this gap, we conducted an iterative design study focusing on how adaptive affective chatbots can better support online learning. Specifically, the study intends to investigate whether affective chatbot interaction can address online learners’ emotional and motivational challenges. In the initial needfinding phase, we identified substantial variation in how online learners perceive and respond emotional and motivational support styles, with no single strategy universally preferred. This led us to shift focus from predefined affective response styles toward adaptive interaction approaches, through which we developed and evaluated three adaptive design alternatives—user-controlled adaptation, AI-driven adaptation, and hybrid co-controlled adaptation—followed by the implementation and evaluation of a refined high-fidelity prototype. We then reflect on how different adaptive interaction strategies shaped learners' experience and discuss the trade-offs involved in designing adaptive conversational agents for online learning.

This work makes the following contributions:
\begin{itemize}
    \item An iterative empirical exploration of how online learners perceive different adaptive affective support strategies in chatbot interaction;
    
    \item A comparative analysis of trade-offs between automation, transparency, and user control in adaptive affective support design;
    
    \item Design implications for affective chatbot systems supporting emotionally challenging online learning contexts.
\end{itemize}

% [OPTIONAL INTRO ADDITIONS]
% - Brief sentence connecting rise of LLM-based educational tools (e.g., ChatGPT in online learning)
% - Short framing on adaptive systems / mixed-initiative interaction
% - Short positioning sentence distinguishing this work from purely pedagogical chatbot research

\section{Related Work}

\subsection{Emotion and Online Learning}

From a socio-cultural constructionist perspective, meaningful learning is closely connected to the social, cultural, and contextual factors surrounding it, emphasizing both cognitive and affective dimensions of the learning process \cite{lund2008wikis,vygotsky1978mind}. In this context, students' emotional states influence their learning experience, either enhancing or hindering their engagement \cite{picard2005evaluating,sheng2010model}. Prior research has also shown that emotional engagement plays a significant role in students' academic performance, motivation, and persistence \cite{yu2022investigating}.

The significance of emotional aspects becomes particularly salient in online learning environments (OLEs), where learners often experience isolation, fluctuating motivation, and limited emotional support structures compared to face-to-face settings. For instance, positive emotional states such as enjoyment and excitement have contributed to sustained engagement in online learning, whereas frustration, boredom, and anxiety can disrupt learning processes and academic outcomes \cite{obergriesser2020students,cho2015self, lobczowski2021socioemotional}. Furthermore, when learners receive emotional support and their needs are acknowledged, they are more likely to maintain self-efficacy and persistence, particularly during cognitively demanding or stressful situations often experienced in online learning environments \cite{miner2020chatbots}.

\subsection{Affective Chatbots in Education}

Recent research on affective and emotionally intelligent chatbots has explored how conversational agents can provide emotional support and improve engagement and trust.Prior work suggests that emotionally aware chatbot responses can reduce miscommunication and increase perceived trustworthiness \cite{portela2017new}. In addition, adopting empathetic tones has been shown to reduce user stress, and matching chatbot personality to user preferences can improve interaction quality , although the effectiveness of different interaction styles varies depending on context, task, and user demographics \cite{smestad2019chatbot, degennaro2020effectiveness}.

In educational contexts, prior studies have shown that learners’ experiences can be improved through personalization and adaptive interaction with chatbots, particularly when chatbots exhibit human-like affective behavior \cite{dmello2014confusion}. Research in adjacent domains, such as mental wellness chatbots, further highlights the potential of emotionally supportive conversational systems as accessible support tools, though concerns related to emotional engagement, contextual appropriateness, and user trust remain important challenges. In addition, empathic chatbots can provide emotional support comparable to certain forms of human interaction, especially in mitigating the effects of social exclusion \cite{degennaro2020effectiveness}. Despite these findings, existing studies have paid less attention to how affective chatbot interactions can be designed to address the specific motivational and emotional needs of online learners.

\section{Needfinding Study}
To understand students’ emotional experiences in online learning and identify design opportunities for effective chatbot support, we conducted a mixed-methods needfinding study from an online forum in graduate course, consisting of a survey, semi-structured interviews, and heuristic evaluations of existing chatbot systems.

\subsection{Online Survey}
We conducted an online survey with participants recruited from a course discussion forum using Likert-scale and short-response items to capture learners' emotional experiences, motivational needs, and preferred chatbot interaction styles. The survey covered four categories: AI usage, affective experiences (e.g., overwhelm, tone comfort), motivational support (e.g., encouragement), and cognitive perceptions (e.g., human-like tone, support style).

Responses from 38 participants indicate that emotional strain is a common aspect of online learning, with moderate levels of overwhelm reported (M = 3.3/5). While learners were generally receptive to empathetic and motivational feedback (M = 3.6--3.8), preferences varied across individuals, with some favoring efficiency over emotional acknowledgment. Support style preferences further reflected this variability, where the Encouraging Coach (37\%) and Calm Mentor (32\%) were most favored, while a substantial portion (26\%) preferred a Direct Problem Solver, suggesting a preference for emotionally aware but also goal-oriented support.

Open-ended responses (n = 18) further indicate that chatbot support is most useful when it combines structured, actionable guidance with light emotional acknowledgment. Participants emphasized responses that are "clear and easy to follow" while also being "natural without fluff", indicating a preference for communication that is concise and action-oriented.
 
\subsection{Semi-Structured Interviews}
To better understand learners’ emotional experiences and expectations for support in online learning, we conducted semi-structured interviews with four learners from diverse academic backgrounds (e.g., MBA and MS programs) who use chatbots as part of their study process. The interviews explored how learners describe their emotional states, what types of support they find helpful during challenging moments, and how they perceive different response styles (e.g., empathetic, motivational, or neutral).

The responses mostly aligned with survey findings on emotional strain and their online learning challenges. However, preferences for support varied across individuals. Some preferred direct, task-oriented responses focused on problem-solving, while others emphasized the importance of empathy and reassurance before receiving guidance. For example, one participant noted that "being extra empathetic would feel fake," while another expressed the need for responses that "acknowledge how stressful it feels" before offering help. Participants also highlighted that mismatched support could be counterproductive: overly empathetic responses were sometimes perceived as inauthentic, while purely task-focused responses were described as dismissive.

\subsection{Heuristic Evaluation}
To identify limitations in existing AI chatbot systems, we conducted a heuristic evaluation of three widely used conversational agents: ChatGPT, Claude, and Microsoft Copilot. Using two simulated scenarios (i.e., frustration and low self-efficacy), we evaluated the systems using heuristics related to system responsiveness, emotional support, motivational guidance, and tone. Although all three systems demonstrated strong cognitive support, their ability to support the emotional and motivational needs of users was inconsistent. Across the three systems, we observed three recurring patterns that informed our subsequent design exploration.

\textbf{Task-Centric Bias and Contextual Forgetting:} Agents often failed to sustain emotional awareness across an interaction. For instance, ChatGPT initially acknowledged user emotions but quickly shifted focus toward task completion without maintaining the user’s prior emotional context.

\textbf{Formulaic Empathy and Cognitive Overload:} In some cases (e.g., Claude), affective responses appeared rigid or formulaic, relying on generic expressions that may feel inauthentic. Directing users immediately toward problem-solving without sufficient emotional acknowledgment may increase cognitive load or reduce perceived support.

\textbf{Prompt-Dependent Motivation:} While some agents (e.g., CoPilot) showed high potential for empathy, the quality of motivational support varied depending on user prompting rather than being consistently integrated into the interaction.

\subsection{Key Findings} 
Across the needfinding process, three key findings emerged regarding the diverse affective and support needs of online learners and the limitations of existing chatbot systems in addressing those needs.

\textbf{(1) Support Preferences are Highly Individualized.}
Learners differ in how they respond to support styles. While some prefer direct, task-oriented guidance, others value empathetic and reassuring responses. This indicates that a single static support style may not effectively accommodate diverse emotional and motivational needs.

\textbf{(2) Emotional Strain and the Need for Actionable Empathy.} Learners frequently experience frustration, isolation, and cognitive overload in online learning. Emotional acknowledgment alone is often perceived as insufficient unless paired with structured, actionable guidance combining light emotional validation with clear, step-by-step assistance.

\textbf{(3) Limitations of Current Systems in Affective Persistence.} The heuristic evaluation revealed that existing chatbots do not consistently sustain affective awareness throughout interactions within the evaluated scenarios, often resulting in a task-centric shift where emotional context is not maintained over time. Affective responses may also appear generic or formulaic, with motivational support varying in quality depending on user prompting rather than being consistently integrated.

\begin{comment}
[These findings indicate that support preferences vary across individuals and contexts, suggesting the need for more adaptive approaches to chatbot interaction.

These findings indicate that learners’ support needs differ not only across individuals but also across situations, suggesting that a single fixed response style may not be sufficient to support diverse learning experiences.]

[These findings highlight gaps in current chatbot design, particularly in balancing cognitive support with consistent and context-sensitive emotional support, motivating the need for more adaptive approaches.]
\end{comment}

\section{Formative Design Process}

\subsection{Design Rationale} 

Findings from our needfinding phase reveal a mismatch between limited chatbot support in existing systems and the diverse emotional and cognitive needs of online learners. The wide variation in preferred support styles indicates that no single static response style can effectively accommodate emotional strain and fluctuating motivation of online learners. Furthermore, existing chatbot systems show limitations in maintaining consistent and context-sensitive affective support across interactions. 

To address this gap, we shifted our focus from optimizing a single response style to exploring how adaptive interaction strategies — integrating different tones (empathetic, motivating, and neutral) with adjustable explanation styles — could be structured as distinct adaptation approaches.

\subsection{Design Alternatives}
To examine the design space of different adaptive strategies, we explored three alternative approaches and designed three different mid-fidelity prototypes in Figma. Each alternative explores a different point on the spectrum between manual control and system-led automation.

\begin{figure*}[t]
    \centering

    \begin{subfigure}[b]{0.32\textwidth}
        \centering
        \includegraphics[width=\linewidth]{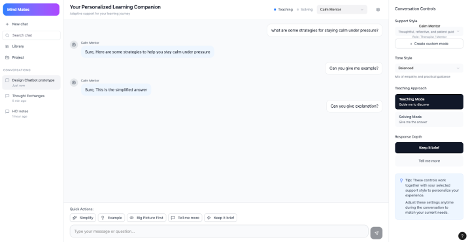}
        \caption{Prototype A: User-Controlled}
        \label{fig:mid_proto_a}
    \end{subfigure}
    \hfill
    \begin{subfigure}[b]{0.32\textwidth}
        \centering
        \includegraphics[width=\linewidth]{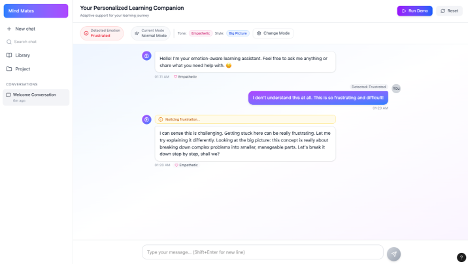}
        \caption{Prototype B: AI-Driven}
        \label{fig:mid_proto_b}
    \end{subfigure}
    \hfill
    \begin{subfigure}[b]{0.32\textwidth}
        \centering
        \includegraphics[width=\linewidth]{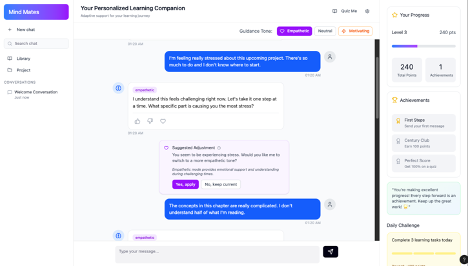}
        \caption{Prototype C: Hybrid}
        \label{fig:mid_proto_c}
    \end{subfigure}

    \caption{Three mid-fidelity design alternatives for adaptive affective support: 
    (a) user-controlled adaptation, (b) AI-driven adaptation, and 
    (c) hybrid adaptation combining system suggestions with user control.}
    \Description{Three chatbot interface screenshots illustrating user-controlled, AI-driven, and hybrid approaches to adaptive affective support.}
    \label{fig:design_alternatives}
\end{figure*}

\textbf{(1) User-Controlled Adaptation.} Based on needfinding results showing large individual differences in preferred tone and support depth, and that learners require different types of assistance at different points, this approach allows learners to manually select or switch support modes depending on their current emotional or cognitive needs. Three primary modes are offered: Calm Mentor (default), Encouraging Coach, and Problem Solver. Additional UI controls allow learners to adjust explanation style on demand, including options to simplify content, request more detail, switch between Teaching Mode and Solving Mode, or toggle between a big-picture overview and an example-first approach. The design goal is to prioritize transparency and predictability, giving learners full control over how the chatbot responds.

\textbf{(2) AI-Driven Adaptation.}
Grounded in needfinding results showing that learners' emotional and cognitive needs fluctuate dynamically, and that existing systems lack consistent affective attunement across interactions, this approach automatically adjusts tone and support style based on inferred user states such as frustration, confusion, or stress. When negativity is detected, the chatbot shifts toward encouraging responses and, if distress persists, activates a forced cooldown or Vent/Rant mode that temporarily deprioritizes task progress in favor of emotional acknowledgment. For cognitive adaptation, the chatbot detects contextual confusion and auto-adjusts explanation complexity, including automatically selecting between a big-picture overview or an example-first approach depending on the inferred learner state. To help learners understand the chatbot's behavior, simple transparency cues are implemented as inline indicators (e.g., "Noticing frustration..."). By minimizing the need for manual interaction, the design is intended to prioritize ease of use.

\textbf{(3) Hybrid Adaptive System.}
Given that some participants prefer not to rely on fully automated systems, while others find manual control burdensome, this approach reflects a co-controlled adaptation model between the user and the chatbot. Combining AI-driven suggestions with user control, this version features an ongoing feedback loop (e.g., thumbs up/down, emotional affinity ratings) and system-initiated check-in points for pacing adjustment, alongside a phased step-by-step plan in which the AI proposes tone adjustments that the user can confirm or override. To support ongoing engagement and motivation, the design incorporates light gamification elements (e.g., points, challenges), effort-focused and specific praise, success recording with reminders, and a "Quiz me" option for active recall. Preference tracking and interface customization allow the system to refine its behavior over time based on accumulated user input. The goal of this hybrid approach is to balance personalization with user control.

\subsection{Formative Evaluation}
To compare how learners perceive different adaptive approaches, we conducted an online study ($n = 25$) in which participants reviewed three mid-fidelity prototypes. Participants evaluated each prototype based on transparency, perceived control, and interface simplicity, followed by cross-comparison questions to compare overall preferences across the three approaches.

The results showed distinct trade-offs across the three approaches. The hybrid approach (Prototype C) received the highest ratings for transparency (M = 4.44) and user control (M = 4.20). This indicates that participants value AI-driven adjustments, combined with the flexibility of user control. In contrast, the AI-driven approach (Prototype B) received the highest score for simplicity (M = 4.44), suggesting that fully automated systems minimize cognitive effort. The user-controlled approach (Prototype A) scored consistently mid-to-high range, which implies that while control was valued, the required effort may have reduced overall preference.

\begin{table}[h]
\centering
\caption{Summary of Perceived Prototype Effectiveness and User Preferences for Actual Study Tasks ($n=25$)}
\label{tab:prototype_preference}
\small
\begin{tabularx}{\columnwidth}{Xcc}
\hline
\textbf{Prototype} & \textbf{\begin{tabular}[c]{@{}c@{}}Most Effective\\ (Count, \%)\end{tabular}} & \textbf{\begin{tabular}[c]{@{}c@{}}Preferred\\ (Count, \%)\end{tabular}} \\ \hline
A (User-Controlled)      & 8 (32\%)  & 6 (24\%)  \\
B (AI-Driven)            & 7 (28\%)  & 8 (32\%)  \\
C (Hybrid)               & 10 (40\%) & 11 (44\%) \\ \hline
\end{tabularx}
\end{table}

Table 1 summarizes participants’ perceptions of prototype effectiveness and their preferences for actual study tasks across the three adaptive approaches. While the hybrid approach was selected most frequently in both categories, the results also suggest that perceptions of effectiveness did not always directly align with actual usage preference across the prototypes. Regarding overall preference for actual study tasks, the hybrid approach was selected most frequently (44\%), followed by Prototype B (32\%) and Prototype A (24\%). Although these differences were not statistically significant (p > 0.05), the qualitative data provided additional insight into the preference for the hybrid approach.

In qualitative responses, a specific trade-off was further revealed. In the manual approach, participants appreciated the transparency and control it provided but found the interface "busy" and noted that constant manual adjustments "interrupted the flow" of studying. While the AI-driven approach was perceived as simple and participants appreciated the low-effort experience, it was also perceived as unpredictable, and some participants reported feeling “not in control”. The hybrid approach was described as collaborative, and participants valued the AI’s ability to responds to emotional states while retaining the final decision to accept or reject suggestions. The primary drawback was the potential "interface clutter" due to the added feedback mechanisms.

In sum, our formative evaluation highlights a clear trade-off between automation, control, and transparency across the three approaches, where increased automation reduced cognitive effort but also limited user control and transparency. Rather than treating automation and control as opposing strategies, the findings suggested that users preferred systems where adaptation remains visible and controllable.

\begin{comment}
    
[The findings suggest that learners do not prefer controlling every detail manually, but prefer systems that balance automation and control while maintaining transparency.

Participants preferred the hybrid approach overall for its balance of transparency and shared control, while the AI-driven approach was valued for its simplicity, and the user-controlled approach provided clarity but required greater effort. 

Automated adaptation lacked transparency, as users were unsure why changes occurred. Manual control was clear, but it introduced interruptions due to the effort required to adjust settings. Although the hybrid interface was effective, it included unnecessary elements that increased cognitive load.]
\end{comment}

\section{Summative Design and Evaluation}

\subsection{Final Design Implementation and Rationale}

The evaluation results from the previous iteration revealed trade-offs between the three approaches rather than a single dominant solution. Based on the findings, we selected the hybrid adaptive alternative as the foundation for our high-fidelity prototype given its advantages in balancing control and automation as revealed in qualitative data, and refined the design to integrate the strengths of each approach while addressing the limitations.

\begin{figure*}[t]
    \centering
    \begin{subfigure}[b]{0.32\textwidth}
        \centering
        \includegraphics[width=\textwidth, height=4cm]{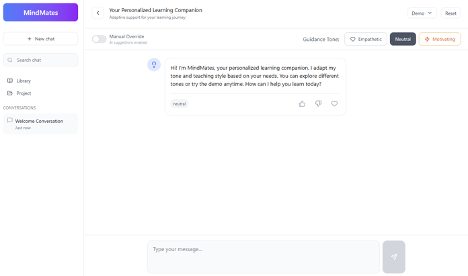}
        \Description{High-fidelity chatbot interface showing a decluttered layout with visible mode indicators.}
        \caption{Decluttered Interface with Mode Indicators.}
        \label{fig:fin_proto_1}
    \end{subfigure}
    \hfill 
    \begin{subfigure}[b]{0.32\textwidth}
        \centering
        \includegraphics[width=\textwidth, height=4cm]{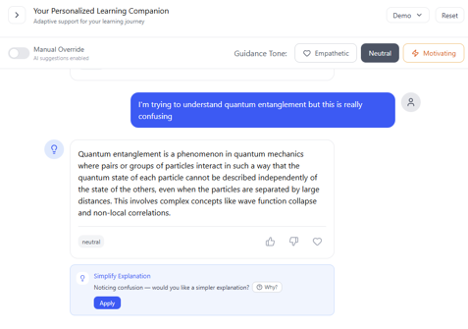}
        \Description{High-fidelity chatbot interface showing contextual inline suggestions during a conversation.}
        \caption{Contextual Inline Suggestions.}
        \label{fig:fin_proto_2}
    \end{subfigure}
    \hfill
    \begin{subfigure}[b]{0.32\textwidth}
        \centering
        \includegraphics[width=\textwidth, height=4cm]{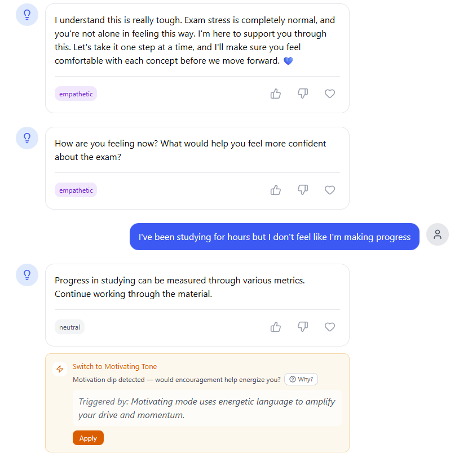}
        \Description{High-fidelity chatbot interface showing a Why label that provides an explanation for an adaptive suggestion.}
        \caption{``Why?'' Transparency Labels.}
        \label{fig:fin_proto_3}
    \end{subfigure}

    \caption{Final high-fidelity hybrid prototype integrating system-initiated affective support with user control and transparency cues.}
    \Description{Three screenshots of the final hybrid prototype illustrating mode indicators, contextual inline suggestions, and transparency labels explaining adaptive suggestions.}
    \label{fig:final_prototype}
\end{figure*}

The final design was implemented as a high-fidelity prototype that balances automation with user control by simplifying interaction flow and making adaptive behavior more visible to users. While three response modes were retained, we refined them to ensure different emotional states are addressed by visually and tonally distinct responses with user approval, using color-coded indicators: neutral (grey) for typical use, empathetic (purple) to regulate stress, and motivating (orange) to aid low self-efficacy. Each chat message is tagged with the active tone mode, making chatbot status continuously visible without requiring additional effort. 

The prototype used predefined, script-driven adaptation, with suggestions and tone changes triggered within fixed demo sequences. To address the perceived unpredictability in the AI-driven approach, the prototype presented contextual inline suggestion cards between messages (e.g., “Switch to Empathetic Mode?”). To minimize disruption to the study flow, learners could accept or dismiss suggestions with a single tap, reducing interaction overhead while preserving control. Each suggestion was accompanied by a toggleable “Why?” button that revealed the rationale behind the adjustment (e.g., \textit{"Empathetic mode uses compassionate language to support you emotionally"}), intended to support transparency and trust through clear communication of the adaptation rationale.

To mitigate the interface complexity, the interface was further simplified by hiding nonessential elements (e.g., navigation menu) using collapsible sidebar to help learners focus solely on the current chat and content. In addition, gamification elements included in the initial hybrid design were removed to address participants' concerns about clutter raised in the formative evaluation. Rather than motivating learners through external rewards, the final design focuses on behavioral transparency and adaptive feedback as the primary engagement mechanism. Figure 1 illustrates the refined high-fidelity hybrid prototype, including the simplified interface, contextual inline suggestions, and transparency labels supporting adaptive interaction visibility.

\subsection{Evaluation Method}
To evaluate the final high-fidelity prototype, we conducted a summative user study with $n=25$ online students. Participants first watched a video walkthrough and interacted with two built-in demo sequences focusing on Cognitive Adjustment (e.g., simplifying complex topics) and Emotional Support (e.g., empathetic/motivating tone shifts). Within these sequences, participants could interact with the adaptation features while observing the corresponding chatbot responses.

After interacting with the prototype, participants completed a survey consisting of nine Likert-scale items and one open-ended question. The evaluation focused on three main aspects of the system: transparency (understanding why adaptations occur), perceived control (ability to influence responses), and cognitive load (ease of interaction). We also assessed the perceived usefulness of adaptive cognitive and emotional support, along with overall preference for the system.

Responses were analyzed using descriptive statistics, and qualitative feedback was coded thematically to identify recurring patterns in user experience.

\subsection{Results and Analysis}
The quantitative results indicated generally positive responses across all measured dimensions. Transparency received the highest rating (M = 4.56, SD = 0.51), followed by perceived control (M = 4.52, SD = 0.59),  and low cognitive load (M = 4.40, SD = 0.71), all significantly above neutral (p < 0.001). These results suggest that the hybrid adaptive chatbot was perceived as understandable to participants while maintaining a sense of control with relatively low effort.

Adaptive features were also perceived as useful. Explanation style adjustments (e.g., simplifying content or providing examples) were rated positively (M = 4.24), while emotional support features (empathetic and motivating tones) received slightly lower but still favorable ratings (M = 4.16). Notably, responses to emotional features showed greater variability compared to usability measures, further reflecting individual differences in how affective support is perceived.

Correlation analysis showed that transparency had a moderate positive relationship with overall helpfulness ($\rho \approx 0.55$, $p < 0.05$), suggesting that clarity of system behavior may play an important role in perceived usefulness. Overall preference favored the adaptive system, with 88\% of participants choosing it over a non-adaptive alternative (p < 0.001).

Qualitative responses revealed similar patterns. Participants favored the chatbot’s ability to adapt to their needs while preserving their sense of control. Many highlighted the balance between guidance and autonomy as a key strength. Compared to earlier prototypes, users also noted improvements in interface simplicity and reduced cognitive load, suggesting that the streamlined design supported more focused interaction.

While most feedback was positive, one participant expressed a general preference for non-adaptive interaction, reflecting a broader skepticism toward AI-mediated emotional responses rather than a specific issue with the system.

\section{Discussion and Reflection}
\begin{comment}
    
\subsection{Key Findings and Interpretation}

Our findings highlight three key patterns in how users interact with adaptive chatbot systems.

First, transparency emerged as a critical factor in perceived usefulness. Participants consistently rated transparency highly, and it showed a stronger relationship with overall helpfulness than other features. This suggests that understanding why the system behaves in a certain way is essential for trust and usability.

Second, users preferred systems that balance automation with user control. Fully automated approaches were seen as simple but unpredictable, while manual control introduced friction. The hybrid approach provided a middle ground, allowing users to remain involved without increasing effort.

Third, responses to emotional support features varied across participants. While some users valued empathetic and motivational tones, others preferred more direct and task-oriented responses. This variability highlights the difficulty of designing universally effective affective interactions.
\end{comment}

\subsection{Reflection on the Design Process}

Our iterative design process revealed that effective affective chatbot support for online learners cannot be reduced to a single optimized response style, but more fundamentally, that the design challenge itself is more structural than stylistic. A central turning point emerged from our needfinding phase, where the wide variation in preferred support styles motivated a shift away from designing a static tone toward exploring adaptive interaction strategies. This pivot was not anticipated at the outset; our initial focus was to investigate how empathetic and motivational chatbot responses could support online learners facing challenges related to isolation, frustration, and engagement identified in prior literature. However, empirical findings made clear that no single style was universally preferred, reframing the problem as one of managing competing design values rather than optimizing a single interaction style. This also aligns with prior studies suggesting that the effectiveness of different affective interaction styles varies depending on context, task, and user demographics [3, 16], extending this observation to the broader challenge of adaptive design

The formative evaluation further surfaced trade-offs that were not initially predictable. Contrary to our expectations, the manual approach — despite scoring well on transparency and control — proved more disruptive to study flow than anticipated, as constant mode switching interrupted participants' engagement. This suggested that what learners appeared to value more was not full control, but rather what might be described as \textit{low-effort agency}: the ability to retain meaningful control without bearing the cognitive cost of managing it. While the AI-driven approach received the highest simplicity ratings, some participants noted that adaptive responses appeared overly automated or emotionally inauthentic, making the system feel less predictable and thus, less preferred. This is in line with prior evidence that perceived trustworthiness is closely tied to the legibility of emotionally aware chatbot responses [12].

The hybrid approach emerged as the most preferred overall, yet it introduced its own challenge: interface clutter from additional feedback mechanisms. One refinement that emerged during iteration involved the removal of gamification elements such as badges and points, which were initially incorporated to encourage motivation and engagement. However, participant feedback suggested that motivational scaffolding may conflict with affective support goals — rather than enhancing the experience, these features introduced additional cognitive load that disrupted learners' focus and interaction flow. These findings surface a possible tension in affective chatbot design, where engagement-oriented features and emotionally supportive interaction goals may not always align, a dynamic that has received limited attention in existing studies on affective chatbots.

Another notable finding concerns the role of transparency in shaping users' receptiveness toward affective AI support. Participants responded more positively when explanations for tone adjustments were made visible through toggleable "Why?" labels. When adaptive behavior was presented transparently rather than as an opaque automated process, participants appeared more receptive to AI-mediated emotional support, suggesting that concerns about affective responses feeling formulaic may be partly attributable to low system legibility rather than adaptation itself. Our findings shed light on the role of system legibility as a critical factor in shaping how learners receive and trust affective AI support.

Importantly, the selection of the hybrid approach as the final design should not be interpreted as an optimal solution, but the most viable trade-off within the scope of this study — one point in a broader design space where persistent tension between automation and control, and between transparency and cognitive load, suggests that affective chatbot design in online learning involves trade-offs that no single iteration can fully resolve across different learning contexts and user preferences.

\subsection{Design Implications}
Our findings suggest several implications for the design of adaptive affective chatbot systems in online learning contexts.

\textbf{Design for Flexible Support Styles.}
Learners differed substantially in how they perceived empathetic, motivational, and task-oriented response styles, suggesting that no single affective tone generalizes across users or situations. Rather than optimizing for a universal response style, future systems may benefit from adaptive mechanisms that allow support styles to remain flexible and context-sensitive over time.

\textbf{Support Low-Effort Agency.}
While learners valued maintaining a sense of control over chatbot behavior, the formative evaluation showed that constant manual configuration disrupted conversational flow and increased interaction burden. Lightweight propose-and-confirm mechanisms, such as inline suggestions with optional confirmation, appeared more effective at balancing user agency with interaction simplicity.

\textbf{Design for Transparent Adaptation.}
Participants responded more positively when adaptive changes were accompanied by visible rationale cues through toggleable "Why?" labels. This suggests that legible system rationale may help users feel more comfortable with AI-mediated affective support without requiring continuous manual oversight.

\textbf{Link Affective Support to Cognitive Guidance.}
Needfinding results suggested that emotional acknowledgment alone was often insufficient unless paired with actionable guidance. Affective support may therefore be more effective when it helps learners transition from emotional overwhelm toward concrete task understanding and problem-solving.

\textbf{Reduce Interaction Overhead.}
In our formative evaluation, gamification elements such as badges and points introduced distraction and additional cognitive load rather than improving engagement. These findings suggest that emotionally supportive learning systems may benefit more from streamlined interaction design than from engagement-oriented features that increase cognitive load.

\begin{comment}

[More broadly, these findings suggest that adaptive affective support may play a complementary role in helping learners navigate emotionally demanding moments in online learning. While this work does not directly evaluate learning outcomes, the observed interaction patterns indicate that balancing emotional support with cognitive guidance may contribute to more manageable and sustainable learning experiences.]
    
\end{comment}

\subsection{Limitations and Future Work}

This study has several limitations. First, both the formative and summative evaluations were conducted as short-term sessions using mid- and high-fidelity prototypes rather than longitudinal deployment in authentic educational settings. As a result, the findings primarily reflect participants’ immediate perceptions of adaptive affective support rather than long-term effects on learning, motivation, or trust calibration over time. Second, the participant pool was relatively small and drawn from a specific academic context, which constrains the generalizability of findings across learner populations with different backgrounds and affective needs. Third, this study did not directly measure learning outcomes. While the observed interaction patterns suggest that adaptive affective support may help learners manage frustration and sustain engagement during challenging tasks, whether these effects translate into measurable improvements in learning-related outcomes remains an open question.

Future work should therefore examine adaptive affective support systems through longitudinal deployments in real online learning environments. In particular, further research may investigate how learners' affective preferences shift over time — both as stable tendencies and in response to situational changes in emotional and cognitive states across different learning moments — and whether transparency mechanisms sustain their effects beyond initial exposure. Low-overhead preference elicitation strategies, such as 
onboarding prompts or in-session conversational cues (e.g., stated preferences and feedback responses), may also offer a path toward personalization while preserving conversational flow. Finally, the tension between engagement-oriented features and affective support goals warrants further investigation as a distinct design challenge.

\begin{comment}
    
\section{Appendices}

If your work needs an appendix, add it before the
``\verb|\end{document}|'' command at the conclusion of your source
document.

Start the appendix with the ``\verb|appendix|'' command:
\begin{verbatim}
  \appendix
\end{verbatim}
and note that in the appendix, sections are lettered, not
numbered. This document has two appendices, demonstrating the section
and subsection identification method.

\end{comment}

\begin{acks}
The author thanks Brandon White, Haemanthi Sree K R, Rodion Dovhan, and Benthara Lakmini Jayathunga for their contributions to data collection and prototype development during the original course project.
\end{acks}

\bibliographystyle{ACM-Reference-Format}
\bibliography{sample-base}

\appendix

\begin{comment}
    
\section{Research Methods}

\subsection{Part One}

Lorem ipsum dolor sit amet, consectetur adipiscing elit. Morbi
malesuada, quam in pulvinar varius, metus nunc fermentum urna, id
sollicitudin purus odio sit amet enim. Aliquam ullamcorper eu ipsum
vel mollis. Curabitur quis dictum nisl. Phasellus vel semper risus, et
lacinia dolor. Integer ultricies commodo sem nec semper.

\subsection{Part Two}

Etiam commodo feugiat nisl pulvinar pellentesque. Etiam auctor sodales
ligula, non varius nibh pulvinar semper. Suspendisse nec lectus non
ipsum convallis congue hendrerit vitae sapien. Donec at laoreet
eros. Vivamus non purus placerat, scelerisque diam eu, cursus
ante. Etiam aliquam tortor auctor efficitur mattis.

\section{Online Resources}

Nam id fermentum dui. Suspendisse sagittis tortor a nulla mollis, in
pulvinar ex pretium. Sed interdum orci quis metus euismod, et sagittis
enim maximus. Vestibulum gravida massa ut felis suscipit
congue. Quisque mattis elit a risus ultrices commodo venenatis eget
dui. Etiam sagittis eleifend elementum.

Nam interdum magna at lectus dignissim, ac dignissim lorem
rhoncus. Maecenas eu arcu ac neque placerat aliquam. Nunc pulvinar
massa et mattis lacinia.
\end{comment}

\end{document}